\documentstyle[11pt,paspconf,epsf]{article}

\begin{document}

\title{Accretion disk in Cyg X-1 in the soft state}

\author{M.\ Gierli\'{n}ski\altaffilmark{1}}
\affil{Jagiellonian University, Astronomical Observatory, Orla 171, 30-244 Krak\'{o}w, Poland}

\author{A.\ A.\ Zdziarski}
\affil{N.\ Copernicus Astronomical Center, Bartycka 18, 00-716 Warszawa, Poland}

\altaffiltext{1}{also N. Copernicus Astronomical Center, Bartycka 18, 00-716 Warszawa, Poland}

\begin{abstract}
We analyze the simultaneous {\it ASCA}/{\it RXTE} observation of Cyg X-1 in the soft state on 1996 May 30. We apply a multi-color disk model with a torque-free boundary condition in the pseudo-Newtonian potential, and point out the main discrepancies between the commonly-used simplified `diskbb' model and our one. We estimate the black hole mass and the accretion rate, and show that during this observation the disk probably did not reach the last stable orbit and the transition from the hard to the soft state was not yet completed.
\end{abstract}



\section{Introduction}

We study the cold disk spectrum of Cyg X-1 in the soft state. The source was observed simultaneously by {\it ASCA} and {\it RXTE} on 1996 May 30. These data were previously analyzed by Dotani et al.\ (1997) and Cui et al.\ (1998). 
Dotani and his collaborators examined the {\it ASCA} data only. They used a multi-color disk blackbody model (`diskbb' in {\sc xspec}, see e.g. Mitsuda et al.\ 1984) and found $kT_{\rm max} = 0.43\pm0.01$ keV and $k_{\rm dbb}R_{\rm in} = 220^{+41}_{-6}$km, where they give the correction factor for `diskbb' model, $k_{\rm dbb} = 25/16 \approx 1.56$. They also applied a general relativistic disk model (`grad' in {\sc xspec}, Hanawa 1989), assuming the disk extends down to the last stable orbit in the Schwarzschild metric, and found a central object mass of $M = 12^{+3}_{-1}$M$_\odot$. Cui et al.\ analysed simultaneous {\it ASCA} and {\it RXTE} observation. Their results concerning the disk parameters are similar to the above ones.

\begin{figure}
\begin{center} \leavevmode 
\hbox{%
\epsfxsize=10cm  
\epsffile{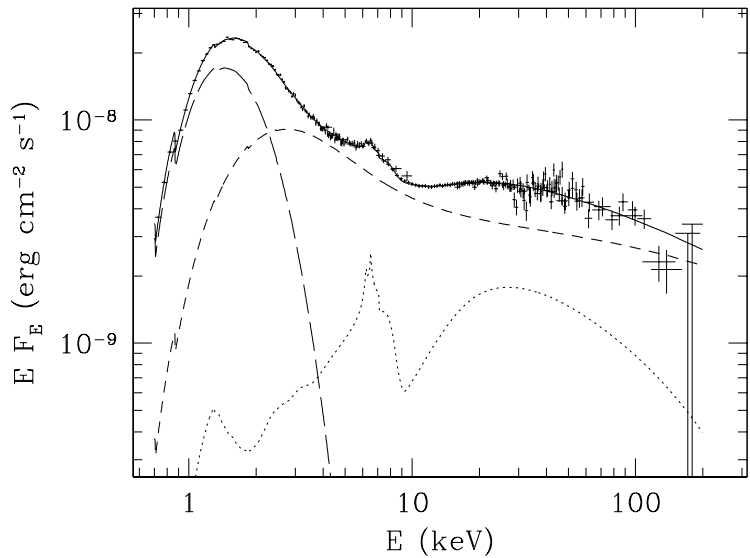}} 
\end{center} 
\caption{The simultaneous {\it ASCA} and {\it RXTE} observation of Cyg X-1 in the soft state on 1996 May 30. The model consists of a non-thermal continuum, Compton reflection from the cold matter and a disk emission (see text for details of the disk model).} \label{fig:spectrum}
\end{figure}

\section{The model}
\label{sec:model}

The soft--state spectrum of Cyg X-1 basically consists of two main components: the cold disk emission below $\sim$3 keV and a power-law continuum extending into $\gamma$-rays (see Figure \ref{fig:spectrum}). We model the disk emission by the multi-color disk blackbody with the torque-free boundary condition in the pseudo-Newtonian potential (see below). For the continuum we use a hybrid thermal/non-thermal model plus Compton reflection from the cold matter (Gierli\'{n}ski et al.\ 1998).
Our cold disk model is as follows. We consider a standard, flat, optically thick accretion disk around a non-rotating black hole of mass $M$. We use the following definitions: $R_g = GM/c^2$, $R_{\rm ms} = 6GM/c^2$, $r = R/R_g$, $\hat{r} = R/R_{\rm ms}$ and $\beta = T_{\rm col}/T_{\rm eff}$. Subscript `in' denotes the inner disk radius. We adopt the pseudo-Newtonian potential (Paczy\'nski \& Wiita 1980),
\begin{equation}
\Phi(R) = -{{GM} \over {R-2R_g}},
\end{equation}
which yields the color temperature distribution along the disk radius,
\begin{equation}
\label{eq:pn_temperature}
T({\hat{r}}) = T_0 \left[3 {{9 {\hat{r}} - 1} \over {{\hat{r}} (3 {\hat{r}} - 1)^3}} \left(1 - {{3 {\hat{r}} - 1} \over {2 {\hat{r}}^{3/2}}}\right)\right]^{1/4},
\end{equation}
where
\begin{equation}
T_0 = \beta \left( {{3 G M \dot M} \over {8 \pi \sigma R_{\rm ms}^3}} \right)^{1/4}.
\end{equation}
The maximum of local temperature is
\begin{equation}
T_{\rm max} = \cases{
0.41 T_0, &${\hat{r}}_{\rm in} < 1.58$,\cr 
T({\hat{r}}_{\rm in}), &otherwise.\cr}
\end{equation}
We create the model spectrum of the accretion disk integrating the observed flux, $F_\nu$, numerically. Our model has three parameters: the inner disk radius, $r_{\rm in} \equiv R_{\rm in}/R_g = 6\hat{r}_{\rm in}$, the maximum temperature, $T_{\rm max}$, and the normalization,
\begin{equation}
K = {{4 \pi} \over {h^3 c^2}} \left(R_{\rm in} \over D\right)^2 {1 \over \beta^4} \cos i,
\end{equation}
from which the absolute value of the inner disk radius, $R_{\rm in}$, can be derived, assuming distance to the source, $D$, and the disk inclination angle, $i$.

\begin{figure}
\begin{center} \leavevmode 
\hbox{%
\epsfxsize=8cm  
\epsffile{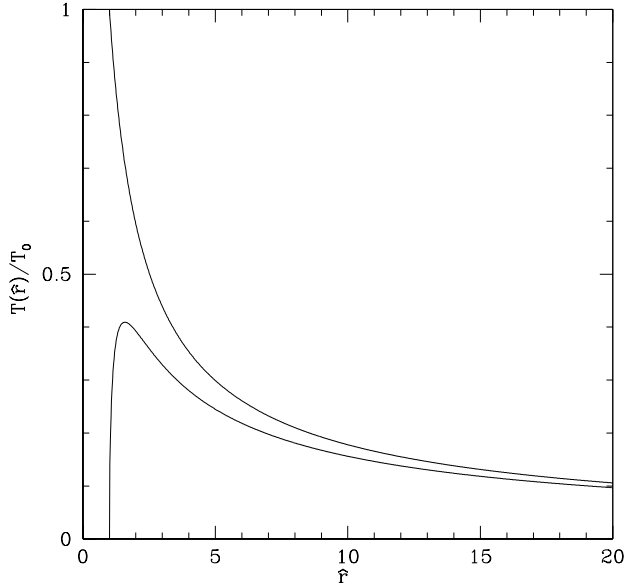}} 
\end{center} 
\caption{Temperature distribution along the disk radius, $\hat{r} \equiv R/R_{\rm ms}$, in a pseudo-Newtonian potential (lower curve) and in a simplified `diskbb' model (upper curve).} \label{fig:disk_temp}
\end{figure}

We point out that the above model differs from the commonly-used `diskbb' model, in which the torque-free boundary condition is neglected and temperature distribution is simplified,
\begin{equation}
\label{eq:diskbb_temperature}
T({\hat{r}}) = T_0 {\hat{r}}^{-3/4}.
\end{equation} 
Discrepancy between the temperature distributions, as shown on Figure \ref{fig:disk_temp}, is crucial when the disk extends down to the last stable orbit. Nevertheless, the integrated spectrum, $F_\nu$, is similar, after some shift in parameters. We find that in the case of $r_{\rm in}$ = 6 our model spectrum for given $R_{\rm in}$ and $T_{\rm max}$ may be approximated by the `diskbb' spectrum, for $R_{\rm in}^{\rm dbb} \approx 2.73R_{\rm in}$ and $T_{\rm max}^{\rm dbb} \approx 1.04R_{\rm max}$.

The two models differ significantly in the accretion efficiency. The total disk luminosity in the pseudo-Newtonian disk (PN) model is
\begin{equation}
L_{\rm PN} = {1 \over 16} \dot{M}c^2,
\end{equation}
giving the accretion efficiency, $\eta_{\rm PN}$ = 1/16 = 0.0625, very close to the exact value in the Schwarzschild metric, $\eta_{\rm S}$ = 1 - (8/9)$^{1/2} \approx$ 0.0572 (e.g.\ Kato, Fukue \& Mineshige 1998), whereas the `diskbb' model yields
\begin{equation}
L_{\rm dbb} = {1 \over 4} \dot{M}c^2,
\end{equation}
with very inaccurate efficiency, $\eta_{\rm diskbb}$ = 1/4. Therefore, the accretion rate calculated from the `diskbb' model is $\sim$4 times less then the actual one.

\section{Disk extending down to the last stable orbit}

In all fits hereafter we assume $D$ = 2.5 kpc, $i$ = 45$^\circ$ and $\beta$ = 1.7.

First, we assume that the optically thick accretion disk extends all the way down to the last stable orbit in the Schwarzschild metric, $r_{\rm in}$ = 6. From the spectral fitting we obtain $kT_{\rm max} = 350\pm10$ eV and $R_{\rm in} = 148^{+11}_{-8}$km, corresponding to a central object mass, $M = 16.7^{+1.2}_{-0.9}$M$_\odot$ and an accretion rate,  $\dot{M} \approx 0.8\times10^{18}$ g s$^{-1}$.

The result obtained from the `diskbb' model (Dotani et al.\ 1997), $k_{\rm dbb}R_{\rm in} = 220^{+41}_{-6}$ km, yields $R_{\rm in} \approx 141$ km, when $k_{\rm dbb}$ = 25/16 is applied. However, as we have shown in the previous section, $k_{\rm dbb} \approx 2.73$, and this result should be rather $R_{\rm in} \approx 81$ km, much less that we have found. The similar value can be computed for the paper by Cui et al.\ (1998). The discrepancy between $R_{\rm in}$ derived from our model and these results comes from the different continuum model assumed. Dotani et al.\ (1997) used simply a broken power law, which fails to approximate properly the Comptonization spectrum at low energies. Cui et al.\ (1998) adopted a thermal Comptonization continuum, which at low energies should agree with our non-thermal model. However, the pure thermal model fails to reproduce the observed high-energy tail and cannot fit the data properly (Gierli\'{n}ski et al.\ 1998). Due to this limitation Cui and his collaborators obtained a very high temperature of seed photons for Comptonization, $kT_s \approx 1$ keV, which, in turn, affected the low-energy part of their model spectrum.

\section{Inner disk edge is farther?}

An interesting result can be obtained, when the relative inner disk radius, $r_{\rm in} \equiv R_{\rm in}/R_g$, is allowed to be a free fit parameter. The shape of the disk spectrum changes slightly with increasing $r_{\rm in}$, and we find these changes large enough to constrain $r_{\rm in}$ by the data. A fit is improved (by $\Delta\chi^2$ = 6.0/574) with the new best-fit value of $r_{\rm in} = 17^{+26}_{-8}$ . The case of the disk extending down to the last stable orbit is {\em outside} the error bars here, suggesting that presumably Cyg X-1 did not finish its transition to the soft state yet. This result is further supported by temporal properties of Cyg X-1. The evolution of power spectra, X-ray time lags and a coherence function (Cui et al.\ 1997) indicate that on 1996 May 30 the source was still on transition from the hard to the soft state, while in June the soft state was established. We note however, that unless all relativistic effects in the strong gravity are properly taken into account, the uncertainties of our disk emission and reflection models do not allow us to make any strong statements.

The other disk parameters found in the $r_{\rm in}$-free model are $kT_{\rm max} =  375^{+3}_{-5}$eV and $R_{\rm in} = 300^{+35}_{-81}$km, yielding $M = 12^{+11}_{-7}$M$_\odot$ and  $\dot{M} \approx 1.2\times10^{18}$ g s$^{-1}$.

\section{Stability of the disk}

To find out whether the cold disk on May 30 did reach the last stable orbit or not, we may try to investigate the stability of the disk. Let us consider a standard Shakura-Sunyaev disk with a fraction $f$ of the gravitational energy dissipated in a corona. This model implies a critical accretion rate, $\dot{m}_{\rm crit}$, above which a radiation-pressure dominated region exists in the disk, which is thought to cause an instability. The critical rate is given by (Svensson \& Zdziarski 1995)
\begin{equation}
\dot{m}_{\rm crit} = 0.37(\alpha m)^{-1/8}(1-f)^{-9/8},
\end{equation}
where $m = M/$M$_\odot$, $\alpha$ is the standard viscosity parameter and the relative accretion rate is defined as $\dot{m} = \dot{M}c^2/L_{\rm Edd}$.

\begin{figure}
\begin{center} \leavevmode 
\hbox{%
\epsfxsize=9cm  
\epsffile{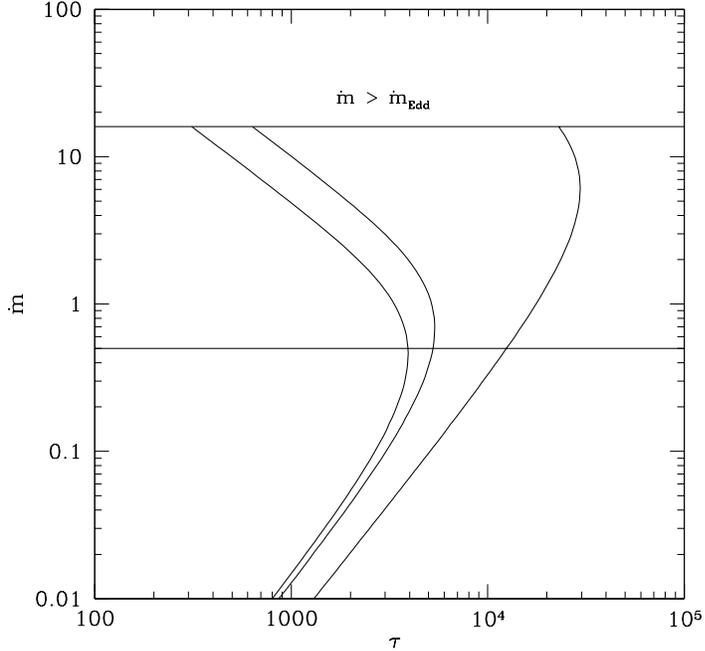}} 
\end{center} 
\caption{The solutions of the optically thick disk with a fraction of accretion power, $f$, dissipated in the corona, for $m$ = 16.7 and $\alpha$ = 0.1, at a radius of the most significant instability $R = 11.4R_{\rm g}$ (Svensson \& Zdziarski 1995). Three curves from the left to the right correspond to $f$ = 0, 0.3 and 0.9. The horizontal line corresponds to the accretion rate of Cyg X-1 in the soft state, $\dot{m} \approx 0.5$.} \label{fig:mdot_tau}
\end{figure}

Now we examine a case of the disk extending down to the last stable orbit, for which we found the mass, $m$ = 16.7, and the accretion rate, $\dot{m}$ = 0.5. From the compactness ratio, $l_h/l_s$ = 0.36 (Gierli\'{n}ski et al.\ 1998), we estimate $f \approx 0.3$. For $\alpha$ = 0.1 the critical rate is $\dot{m}_{\rm crit} \approx 0.52$. The actual accretion rate locates the soft state of Cyg X-1 at the end of a stable branch of the disk solution. Figure \ref{fig:mdot_tau} shows the solutions of the optically thick accretion disk with the corona for $f$ = 0, 0.3 and 0.9.

On the other hand, if we drop the assumption of the disk extending down to the last stable orbit, we find $m$ = 12 and $\dot{m}_{\rm crit} \approx 0.54$. The actual total accretion rate, $\dot{m} \approx 1$, is well above the critical one. The inner part of the disk is thus unstable, which, in fact, is consistent with the truncation in the disk below $r_{\rm in} \sim 17$, we observe. Instead of a cold disk, an advective, optically thin, hot solution might exists in this region. The accretion below 17 $R_g$ dissipates about 0.3 of a total power released in the disk, but due to advection the luminosity of the inner flow might be considerably diminished. Therefore, its thermal emission would be difficult to disentangle from the observed thermal/non-thermal spectrum, and its existence hard to confirm or to rule-out.

\end{document}